\documentclass[epj]{svjour}

\usepackage{latexsym}
\usepackage{graphicx}
\usepackage{fancyhdr}
\def\makeheadbox{{
 }}
\def\mytitle{My title} 
\def\myauthors{My name}  
\def\mytype{My type of session}
\def\mysession{My session}

\def\mytitle{Leptonic Decays at BABAR} 
\def\myauthors{Guglielmo De Nardo}    
\def\mytype{Contributed Talk}    
\def\mysession{Flavor Physics}

\usepackage{bm}
\usepackage{cite}

\usepackage{relsize}
\def\babar{\mbox{\slshape B\kern-0.1em{\smaller A}\kern-0.1em
    B\kern-0.1em{\smaller A\kern-0.2em R}}}

\RequirePackage{xspace}

\def\epem       {\ensuremath{e^+e^-}\xspace}

\def\taup       {\ensuremath{\tau^+}\xspace}

\def\nub        {\ensuremath{\overline{\nu}}\xspace}

\def\nub        {\ensuremath{\overline{\nu}}\xspace}

\def\uubar {\ensuremath{u\overline u}\xspace}

\def\ddbar {\ensuremath{d\overline d}\xspace}

\def\ssbar {\ensuremath{s\overline s}\xspace}

\def\piz   {\ensuremath{\pi^0}\xspace}

\def\Kbar  {\kern 0.2em\overline{\kern -0.2em K}{}\xspace}

\def\Kz    {\ensuremath{K^0}\xspace}
\def\Kzb   {\ensuremath{\Kbar^0}\xspace}
\def\KzKzb {\ensuremath{\Kz \kern -0.16em \Kzb}\xspace}
\def\Kp    {\ensuremath{K^+}\xspace}
\def\Km    {\ensuremath{K^-}\xspace}

\def\KpKm  {\ensuremath{\Kp \kern -0.16em \Km}\xspace}
\def\KS    {\ensuremath{K^0_{\scriptscriptstyle S}}\xspace}

\def\Dbar    {\kern 0.2em\overline{\kern -0.2em D}{}\xspace}

\def\Dz      {\ensuremath{D^0}\xspace}
\def\Dzb     {\ensuremath{\Dbar^0}\xspace}
\def\DzDzb   {\ensuremath{\Dz {\kern -0.16em \Dzb}}\xspace}
\def\Dp      {\ensuremath{D^+}\xspace}
\def\Dm      {\ensuremath{D^-}\xspace}

\def\DpDm    {\ensuremath{\Dp {\kern -0.16em \Dm}}\xspace}

\def\Dstarz  {\ensuremath{D^{*0}}\xspace}

\def\B       {\ensuremath{B}\xspace}
\def\Bbar    {\kern 0.18em\overline{\kern -0.18em B}{}\xspace}

\def\Bz      {\ensuremath{B^0}\xspace}
\def\Bzb     {\ensuremath{\Bbar^0}\xspace}
\def\BzBzb   {\ensuremath{\Bz {\kern -0.16em \Bzb}}\xspace}
\def\Bu      {\ensuremath{B^+}\xspace}
\def\Bub     {\ensuremath{B^-}\xspace}

\def\BpBm    {\ensuremath{\Bu {\kern -0.16em \Bub}}\xspace}

\def\BorBbar    {\kern 0.18em\optbar{\kern -0.18em B}{}\xspace}
\def\DorDbar    {\kern 0.18em\optbar{\kern -0.18em D}{}\xspace}
\def\KorKbar    {\kern 0.18em\optbar{\kern -0.18em K}{}\xspace}

\mathchardef\Upsilon="7107
\def\Y#1S{\ensuremath{\Upsilon{(#1S)}}\xspace}

\def\FourS {\Y4S}

\mathchardef\Deltares="7101
\mathchardef\Xi="7104
\mathchardef\Lambda="7103
\mathchardef\Sigma="7106
\mathchardef\Omega="710A

\def\Deltabar{\kern 0.25em\overline{\kern -0.25em \Deltares}{}\xspace}
\def\Lbar{\kern 0.2em\overline{\kern -0.2em\Lambda\kern 0.05em}\kern-0.05em{}\xspace}
\def\Sigbar{\kern 0.2em\overline{\kern -0.2em \Sigma}{}\xspace}
\def\Xibar{\kern 0.2em\overline{\kern -0.2em \Xi}{}\xspace}
\def\Obar{\kern 0.2em\overline{\kern -0.2em \Omega}{}\xspace}
\def\Nbar{\kern 0.2em\overline{\kern -0.2em N}{}\xspace}
\def\Xb{\kern 0.2em\overline{\kern -0.2em X}{}\xspace}

\def\mes        {\mbox{$m_{\rm ES}$}\xspace}

\newcommand{\tev}{\ensuremath{\mathrm{\,Te\kern -0.1em V}}\xspace}
\newcommand{\gev}{\ensuremath{\mathrm{\,Ge\kern -0.1em V}}\xspace}
\newcommand{\mev}{\ensuremath{\mathrm{\,Me\kern -0.1em V}}\xspace}
\newcommand{\kev}{\ensuremath{\mathrm{\,ke\kern -0.1em V}}\xspace}
\newcommand{\ev}{\ensuremath{\mathrm{\,e\kern -0.1em V}}\xspace}
\newcommand{\gevc}{\ensuremath{{\mathrm{\,Ge\kern -0.1em V\!/}c}}\xspace}
\newcommand{\mevc}{\ensuremath{{\mathrm{\,Me\kern -0.1em V\!/}c}}\xspace}
\newcommand{\gevcc}{\ensuremath{{\mathrm{\,Ge\kern -0.1em V\!/}c^2}}\xspace}
\newcommand{\mevcc}{\ensuremath{{\mathrm{\,Me\kern -0.1em V\!/}c^2}}\xspace}

\def\cm   {\ensuremath{{\rm \,cm}}\xspace}

\def\mus  {\ensuremath{\rm \,\mus}\xspace}

\def\mus        {\ensuremath{\,\mu{\rm s}}\xspace}    

\def\ra                 {\ensuremath{\rightarrow}\xspace}
\def\to                 {\ensuremath{\rightarrow}\xspace}

\def\pep2{PEP-II}

\def\gsim{{~\raise.15em\hbox{$>$}\kern-.85em
          \lower.35em\hbox{$\sim$}~}\xspace}
\def\lsim{{~\raise.15em\hbox{$<$}\kern-.85em
          \lower.35em\hbox{$\sim$}~}\xspace}

\def\Vub  {\ensuremath{|V_{ub}|}\xspace}

\xspace

\def\jetset74   {\mbox{\tt Jetset \hspace{-0.5em}7.\hspace{-0.2em}4}\xspace}

\def\Vub {\ensuremath{V_{ub}}}

\def\btn {\ensuremath{B^{+} \to \tau^{+} \nu}\xspace}

\def\btodx {\ensuremath{\Bub \to D^{(*)0} X^-}}
\def\btodlnux {\ensuremath{\Bub \to \Dz \ell^{-} \bar{\nu}_{\ell} X}}

\def\eextra {\ensuremath{E_{\mathrm{extra}}}\xspace}

\def\tautoenunu {\ensuremath {\tau^+ \to e^+ \nu \nub}}

\def\tautomununu {\ensuremath {\tau^+ \to \mu^+ \nu \nub}}

\def\tautopinu {\ensuremath {\tau^+ \to \pi^+ \nub}}

\def\tautopipiznu {\ensuremath {\tau^+ \to \pi^+ \pi^{0} \nub}}
\def\tautopipiz {\ensuremath {\tau^+ \to \pi^+ \pi^{0} \nub}}

\def\pipiz {\ensuremath { \pi^+ \pi^{0} }}

\begin{document}
\title{Leptonic Decays at BABAR}
\author{Guglielmo De Nardo\inst{1}
\thanks{\emph{Email:} denardo@na.infn.it}%
for the BABAR collaboration
}                     
%
%
\institute{University of Napoli Federico II
INFN Sezione di Napoli }
%
\date{}
\abstract{
We present recent results on leptonic B decays using data collected by the
BaBar detector at the PEP-II asymmetric-energy e+e- collider at the
Stanford Linear Accelerator Center.  We report
searches for the \btn decay based on two statistically independent data samples.
} 
\maketitle
%

\section{Introduction}
\label{intro}
The purely leptonic decay $\btn$~\cite{cc} is 
sensitive to the product of the $B$ meson decay constant $f_{B}$, 
and the absolute value of Cabibbo-Kobayashi-Maskawa matrix element \mbox{$\Vub$~\cite{c,km}}. 
In the Standard Model (SM), the decay 
proceeds via quark annihilation into a $W^{+}$ boson, with a
branching fraction given by:
\begin{equation}
\label{eqn:br}
\mathcal{B}(B^{+} \rightarrow {\taup} \nu)= 
\frac{G_{F}^{2} m^{}_{B}  m_{\tau}^{2}}{8\pi}
\left[1 - \frac{m_{\tau}^{2}}{m_{B}^{2}}\right]^{2} 
\tau_{\Bu} f_{B}^{2} |\Vub|^{2},
\end{equation}
where  
$G_F$ is the Fermi constant,
$\tau_{\Bu}$ is the $\Bu$ lifetime, and
$m^{}_{B}$ and $m_{\tau}$ are the $\Bu$ meson and $\tau$ lepton masses.

The process \btn is also  sensitive to extensions of the SM. 
For instance, in two-Higgs doublet models~\cite{higgs} and in the MSSM~\cite{Isidori2006pk,Akeroyd:2003zr}
 it could be mediated by charged Higgs bosons.
The branching fraction measurement can therefore also be used to constrain
the parameter space of extensions to the SM.
The Belle Collaboration has reported evidence from a search for this decay and the branching fraction was measured to be
$(1.79^{+0.56}_{-0.49}(\mbox{stat.}) ^{+0.46}_{-0.51}(\mbox{syst.})) \times 10^{-4}$~\cite{belle}.
We present here two results from the \babar\ collaboration 
using a sample of $383 \times 10^6$ $\FourS\to\B\Bbar$ decays, 
based on the reconstruction of a semileptonic 
~\cite{taunusemilep} and of an hadronic~\cite{taunuhad}  \textit{B} decay on the tag side.
\section{Tag $B$ Reconstruction}
In the hadronic tags analysis, the tag $B$ candidate is reconstructed 
in hadronic $B$ decay modes 
\btodx, where $X^-$ denotes a system of
charged and neutral hadrons with total charge $-1$
composed of $n_1\pi^{\pm}$, $n_2K^{\pm}$, $n_3\KS$,  $n_4\piz$, where $n_1 + n_2 \leq
5$,  $n_3  \leq  2$,  and  $n_4  \leq  2$.
We  reconstruct $\Dstarz \ra \Dz\piz, \Dz\gamma$; 
$\Dz\ra K^-\pi^+$, $K^-\pi^+\piz$, $K^-\pi^+\pi^-\pi^+$,  $\KS\pi^+\pi^-$ and  $\KS \ra \pi^+\pi^-$. 
Tag $B$ candidates are required to be kinematically consistent with
decay from an \FourS using
the beam energy-substituted mass $\mes = \sqrt{s/4 -
\vec{p}^{\,2}_B}$ and the energy difference 
$\Delta E = E_B - \sqrt{s}/2$. Here $\sqrt{s}$ is the total
energy in the \FourS center-of-mass (CM) frame, and $\vec{p}_B$ and $E_B$
denote, respectively, the momentum and energy of the tag $B$ candidate in the CM
frame.  The purity ${\cal P}$ 
of each reconstructed $B$ decay mode
is estimated, using on-resonance data, 
as the ratio of the number of peaking events with \mes$~>~
5.27$~\gevcc to the total number of events in the same range.
If multiple tag $B$ candidates are reconstructed, the one with the highest purity ${\cal P}$ is selected.
If more than one candidate with the same purity is reconstructed, the one with the lowest value of
$|\Delta E|$ is selected. 
The set of decay modes used is defined by the requirement that the purity
of the resulting sample is not less than 30\%.

The background consists of $\epem\ra q\bar q$ events and
other $\FourS\to\BzBzb$ or \BpBm decays
in which the tag $B$ candidate is mis-reconstructed using particles 
coming from both $B$ mesons in the event.
To  reduce the $\epem\ra q\bar q$ background,  we require 
$|\cos{\theta_{TB}^*}|<0.9$, where $\theta_{TB}^*$ is 
the angle in the CM frame between the thrust
axis~\cite{thrust} of the tag $B$ candidate and the thrust axis of the 
remaining reconstructed charged and neutral candidates.

\begin{figure}[!thb]
  \begin{center}
    \includegraphics[width=0.90\linewidth]{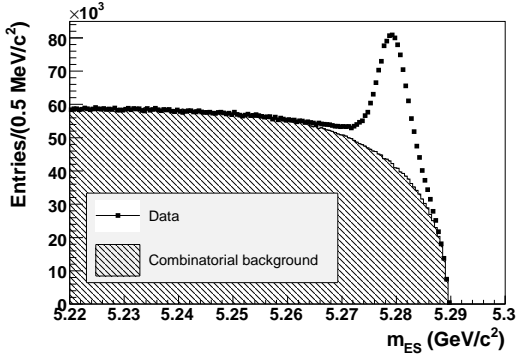}

    \caption{Distribution of the energy substituted mass, \mes, 
      of the tag $B$ candidates in data. The combinatorial background is overlaid. 
    } 
    \label{fig:mesfit}
  \end{center}
\end{figure}

In order to determine the number of correctly reconstructed $\Bu$ decays,
the background events are classified in four  categories: $\epem\ra c\bar c$;
$\epem\ra\uubar,\ddbar,\ssbar$; \FourS\ra\BzBzb; and \FourS\ra\BpBm. 
The \mes shapes of these background distributions are taken from MC simulation.
The normalization of the  $\epem\ra c\bar c$ and $\epem\ra \uubar,\;\ddbar,\;\ssbar$ 
backgrounds is taken from off-resonance data, scaled by the luminosity and corrected for the 
different selection efficiencies evaluated with the MC.
The normalization of the \BzBzb, \BpBm components are 
obtained by means of  a $\chi^2$ fit to the \mes distribution in 
the data sideband region ($5.22\gevcc<\mes<5.26\gevcc$). 
The number of background events in the signal region ($\mes>5.27\gevcc$) 
is extrapolated from the fit and subtracted from the data. 
We estimate the total number of tagged $B$'s in the data to be
$N_{B} = (5.92 \pm 0.11\textrm{(stat)}) \times 10^{5}$.
Figure \ref{fig:mesfit} shows the tag $B$ candidate \mes\ distribution,
with the combinatorial background, 
estimated as the sum of the four components described above, overlaid.  

In the semileptonic tag analysis,
the tag $B$ is reconstructed in semileptonic $B$ decay modes 
\btodlnux, where $\ell$ is either an electron or a muon, and $X$ can be either nothing, 
a $\piz$ or a photon.
Events where the best tag candidate is consistent with neutral $B$ decay are rejected.
The identified electron or muon in the $\Dz\ell$ candidates are required to 
have  momentum above $0.8\gevc$ in the $e^+e^-$ center-of-mass (CM) frame.
The flight direction of the $\Dz$ is required to intersect with
the lepton track.
We reconstruct the $\Dz$ candidates in four decay modes:
$K^{-}\pi^{+}$, $K^{-}\pi^{+}\pi^{-}\pi^{+}$, $K^{-}\pi^{+}\pi^{0}$, and
$\KS\pi^{+}\pi^{-}$, only considering $\KS$ candidates decaying to charged pions. 
The $\piz$ candidates are required to have invariant masses between
0.115 and 0.150$\gevcc$ and the photon daughter candidates of the $\piz$ must 
have a minimum laboratory energy of 30$\mev$ and have shower shapes consistent with 
electromagnetic showers. 
The mass of the reconstructed $\Dz$ candidates in the
$K^{-}\pi^{+}$, $K^{-}\pi^{+}\pi^{-}\pi^{+}$, and $\KS\pi^{+}\pi^{-}$
modes is required to be within 20$\mevcc$ of the nominal mass 
\cite{pdg2004}, while in the $K^{-}\pi^{+}\pi^{0}$ decay mode 
the mass is required to be within 35$\mevcc$ of the nominal mass. 
Furthermore, the sum of the charges of all the particles 
in the event must be equal to zero.

We calculate the cosine of the angle between the $\Dz\ell$ candidate
and the $B$ meson as
\begin{equation}
\cos\theta_{B-\Dz\ell} = \frac{2 E_{B} E_{\Dz\ell} - m_{B}^{2} - 
m_{\Dz\ell}^{2}}{2|\vec{p}_{B}||\vec{p}_{\Dz\ell}|},
\label{eqn:cosby}
\end{equation}
where ($E_{\Dz\ell}$, $\vec{p}_{\Dz\ell}$) is the 
four-momentum of the $\Dz\ell$ candidate in the CM frame, and $m_{\Dz\ell}$ and $m_{B}$ 
are the invariant masse of the $\Dz\ell$ candidate and the $\Bu$ meson
nominal mass~\cite{pdg2004}, respectively. 
We expect $\cos\theta_{B-\Dz\ell}$ for correctly reconstructed tag B candidates 
to be in the range [$-1,1$], whereas combinatorial backgrounds
can have values outside this range. 
We select events with $-2.0 < \cos\theta_{B-\Dz\ell} < 1.1$, 
If multiple tag are reconstructed the $\Dz\ell$ candidate
with the largest probability of originating from a single vertex is selected.

From signal MC we estimate the tag reconstruction efficiency to be 
$(6.64\pm0.03)\times 10^{-3}$, where the 
error is due to the statistics of the signal MC sample. This corresponds to a tag $B$
yield of $(2.54\pm0.03)\times 10^6$.

\section{Selection of Signal Events}
After the reconstruction of the tag $B$ meson,
the rest of the event (recoil) is examined for \btn decays.
We require the presence of only one well-reconstructed charged track (signal track) 
with charge opposite to that of the tag $B$. The signal track is required to have at least
12 hits in the drift chamber, momentum transverse to the 
beam axis, $p_{T}$, greater than 0.1$\gevc$, and
the point of closest approach to the interaction point
less than 10\cm along the beam axis and less than 1.5\cm transverse 
to it.

The $\tau$ lepton is identified in four decay modes
constituting approximately 71\% of the total $\tau$ decay width: $\tautoenunu$, $\tautomununu$, $\tautopinu$, and \tautopipiznu .
Particle identification criteria on the signal track are used to separate the four 
categories. 

The  \tautopipiznu\ sample is obtained by associating 
the signal track, identified  as pion, with a $\pi^0$ reconstructed from a pair of neutral 
clusters with invariant mass 
between 0.115 and 0.155 \gevcc. In the hadronic tag analysis 
and \pipiz\ energy is required to be greater than 250 \mev and
in case of multiple  candidates, the one with largest
center-of-mass momentum $p^*_{\pi^+\piz}$ is chosen. In the semileptonic tag analysis
the energy of the neutral clusters is required to be greater than 50 \mev. 

The selection is further refined with additional requirements exploiting the kinematics
of the signal, including cuts on the momentum of visible \taup\ decay products, the missing momentum,
the missing mass, the invariant mass of the \pipiz\ candidate, charged tracks and
\piz candidates multiplicities.
The semileptonic tag analysis also includes a veto on signal of $K^0_L$ in the calorimeter
and in the muon detector.

We optimize the selection by maximizing $s/\sqrt{s+b}$ using the
MC simulation, where $b$ is the expected background  
and $s$ is the expected number of signal events 
in the hypothesis of a branching fraction 
of $1 \times 10^{-4}$. The optimization is performed separately for each $\tau$ decay mode and 
with all the cuts applied simultaneously in order to take into account any correlations  
among the discriminating variables.
The semileptonic tag analysis uses the PRIM algorithm~\cite{prim1999} to find
the optimal set of cuts.

For both the analyses, the most powerful 
discriminating variable is \eextra defined as the sum of the energies 
of the neutral clusters not associated
with the tag $B$ or with the signal $\pi^0$ from the \tautopipiz\ mode,
and passing a minimum energy requirement. 
In the hadronic tag analysis the required energy depends 
on the selected signal mode and on the calorimeter region involved and varies from
50 to 70 \mev. In the semileptonic tag analysis, the minimum energy requirement is
fixed to 20 \mev, and the energy of charged tracks not associated with the tag $B$ or
the signal candidate are included in the \eextra computation.
Signal events peak at low \eextra values, 
whereas background events, which contain additional sources 
of neutral clusters, are distributed toward higher \eextra values.

The total selection efficiency is estimated from signal MC to be
$\varepsilon = (9.8 \pm 0.3) \%$, and $\varepsilon = (12.7 \pm 0.2) \%$
for the hadronic tag analysis and the semileptonic tag analysis, respectively.

\section{Background yield}
In the hadronic tag analysis, we define a sideband region 
$ 0.4 \gev< \eextra < 2.4 \gev$ and a $\tau$ mode-dependent signal
regions.
We perform an extended unbinned maximum likelihood fit to
the \mes distribution in the  $\eextra$ data sideband region of the final sample.
For the peaking component of the background we use a probability density function (PDF) 
which is  a Gaussian function joined
to an exponential tail (Crystal Ball function)~\cite{crystalball}. As a PDF for the non-peaking component,
we use a phase space motivated threshold function (ARGUS function)~\cite{arguspdf}.
From this fit, we determine a peaking yield  $N_{pk}^{\rm{side},\rm{data}}$ 
and signal shape parameters, to be used in later fits. 
The same procedure is applied to \BpBm MC events which 
pass the final selection to determine the peaking yield $N_{pk}^{\rm{side},\rm{MC}}$.
To determine the MC peaking yield in the \eextra signal region 
$N_{pk}^{\rm{sig},\rm{MC}}$, 
we fit \mes in the \eextra signal region of the \BpBm MC sample
with the Crystal Ball parameters fixed to the values determined
in the \eextra sideband fits described above.
Analogously, we fit the  \mes distribution of data in the \eextra signal region 
to extract the combinatorial  background $n_{\rm{comb}}$,
evaluated as the integral of the ARGUS shaped component 
in the $\mes > 5.27 \gevcc$ region.
The total expected background in the signal region is determined as
\begin{equation}
b = \frac{ N_{pk}^{\rm{sig},\rm{MC}}} {N_{pk}^{\rm{side},\rm{MC}}} \times N_{pk}^{\rm{side},\rm{data}} + n_{\rm{comb}}.
\label{eq:bgformula}
\end{equation}
In the semileptonic tag analysis, the sideband region is defined by
$\eextra>0.5\gev$, and the signal regions in $\eextra$ are signal mode-dependent.
Using the number of events in the sideband ($N_{\mbox{\scriptsize{MC,sb}}}$) 
and signal ($N_{\mbox{\scriptsize{MC,sig}}}$) regions from MC simulation and 
the number of data events in the sidebands $N_{\mbox{\scriptsize{data,sb}}}$,
we estimate the number of expected background events in the signal region
in data $N_{\mbox{\scriptsize{exp,sig}}}$

\begin{eqnarray}
N_{\mbox{\scriptsize{exp,sig}}} & = & N_{\mbox{\scriptsize{data,sb}}} \cdot \frac{N_{\mbox{\scriptsize{MC,sig}}}}{N_{\mbox{\scriptsize{MC,sb}}}}.
\end{eqnarray}

The background estimate is validated using sidebands in the $\Dz$ mass distribution.

\section{Results}
\begin{figure}[htb]
\includegraphics[width=0.9\linewidth]{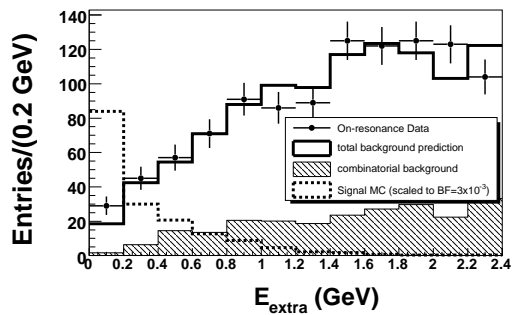} \\
\caption{$\eextra$ distribution after all selection criteria 
have been applied. 
The on-resonance data (black dots) distribution is compared with the total background prediction
(continuous histogram). The hatched histrogram represents the combinatorial background component.
$B^+\to\tau^+\nu$ signal MC (dashed histogram), normalized to a branching fraction of $3\times10^{-3}$ for illustrative purposes, is shown for comparison. }
\label{fig:eextra}
\end{figure}

We measure the yield of events in each decay mode in on-resonance data. 
Table~\ref{tab:unblres} reports, for the hadronic tag analysis, the
number of observed events together with the expected number of 
background events, for each $\tau$ decay mode.
Figure \ref{fig:eextra} shows the \eextra distribution for data and
expected background at the end of the selection. The signal MC,
normalized to a branching fraction of $3\times10^{-3}$ for illustrative purposes, 
is overlaid for comparison.
The \eextra distribution is also plotted separately for each $\tau$ decay mode.  

\begin{table}
\caption{
Observed number of on-resonance data events in the signal region compared
 with the number of expected background events, for the hadronic tag analysis.}
\label{tab:unblres}
\begin{tabular}{lcc} \hline 
$\tau$ decay mode   & Expected background  &  Observed   \\ 
\hline \hline
\tautoenunu	  & 1.47  $\pm$ 1.37   & 4  \\ 
\tautomununu      & 1.78  $\pm$ 0.97   & 5  \\ 
\tautopinu        & 6.79  $\pm$ 2.11   & 10 \\ 
\tautopipiznu     & 4.23  $\pm$ 1.39   & 5  \\ 
\hline
All modes    & 14.27 $\pm$  3.03 & 24  \\ \hline \hline
\end{tabular}
\end{table}

\begin{figure}[htb]
\begin{center}
\includegraphics[width=.35\textwidth]{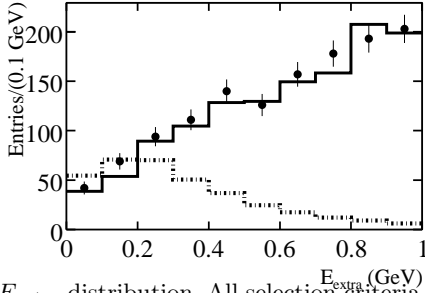}
\end{center}
\vspace{-0.8cm}
\caption{$\eextra$ distribution. All selection criteria 
have been applied and all signal modes combined. 
Background MC (solid histogram) has been normalized to the luminosity 
of the on-resonance data (black dots), and then
additionally scaled according
to the ratio of predicted background from data and MC.
$\btn$ signal MC (dotted histogram) is normalized to a branching
fraction of $10^{-3}$ and shown for comparison.}
\label{fig:eextra_allcuts_ALL}
\end{figure}

Table~\ref{tab:unblind-result} lists, for the semileptonic tag analysis, the
number of observed events in on-resonance data in the signal region,
together with the expected number of background events in the 
signal region.
Figure~\ref{fig:eextra_allcuts_ALL} shows the $\eextra$
distribution for all data and MC in the signal region, with signal MC shown for
comparison.

\begin{table}[hbt]
\centering
\caption{\label{tab:unblind-result}
Observed number of on-resonance data events in the signal region are shown, 
together with number of expected background events,
for the semileptonic tag analysis.}
\begin{tabular}{lcc} \hline \hline
$\tau$            & Expected background  &  Observed  \\ \hline
$\tautoenunu$     & 44.3  $\pm$ 5.2   & 59  \\ 
$\tautomununu$    & 39.8  $\pm$ 4.4   & 43  \\ 
$\tautopinu$      & 120.3 $\pm$ 10.2  & 125  \\ 
$\tautopipiznu$   & 17.3  $\pm$ 3.3   & 18  \\ 
\hline
All modes    & 221.7 $\pm$ 12.7  & 245  \\ \hline \hline
\end{tabular}
\end{table}

In both the analyses, we combine the results on the observed number of events  $n_i$ and on the expected background
$b_i$  from each of the four signal decay modes ($n_{ch}$) using the estimator
\mbox{$Q = {\cal L}(s+b)/{\cal L}(b)$},
where ${\cal L}(s+b)$ and ${\cal L}(b)$ are the
likelihood functions for signal plus background and background-only
hypotheses, respectively:
\begin{equation}
  {\cal L}(s+b) \equiv
  \prod_{i=1}^{n_{ch}}\frac{e^{-(s_i+b_i)}(s_i+b_i)^{n_i}}{n_i!},
        \;
  {\cal L}(b)   \equiv
  \prod_{i=1}^{n_{ch}}\frac{e^{-b_i}b_i^{n_i}}{n_i!}.
  \label{eq:lb}
\end{equation}
The  estimated number of signal candidates $s_i$ in data, for each decay mode, is related to the 
\btn branching fraction by:
\begin{equation}
s_i = \frac {   \varepsilon^{\rm{tag}}_{\rm{sig}} } { \varepsilon^{\rm{tag}}_{B} }
      N^{\rm{tag}}_{\Bu}\varepsilon_i  \mathcal{B}(\btn),
\end{equation}
where $N^{\rm{tag}}_{\Bu}$ is the number of tag $\Bu$ mesons 
correctly reconstructed, 
$ \varepsilon^{\rm{tag}}_{B}$ and $\varepsilon^{\rm{tag}}_{\rm{sig}}$ are the tag $B$ efficiencies in 
generic $B\bar{B}$ and signal events respectively, 
and $\varepsilon_i$ are the signal selection efficiencies for each channel,
including the \taup\ branching fractions. 
In the hadronic tag analysis, we find the ratio from MC simulation to be
$\varepsilon^{\rm{tag}}_{\rm{sig}} / \varepsilon^{\rm{tag}}_{B}  = 0.939\pm0.007\textrm{(stat.)}$.

We estimate the branching fraction
(including statistical uncertainty and uncertainty from the background~\cite{giunti}) 
by scanning over signal branching fraction hypotheses 
and computing the value of $\mathcal{L}(s+b)/\mathcal{L}(b)$ for each
hypothesis. The branching fraction is the hypothesis which minimizes the likelihood ratio 
$-2 \ln \textrm{Q}= -2 \ln(\mathcal{L}(s+b)/\mathcal{L}(b))$,
and we determine the  statistical uncertainty  by finding the points on the likelihood scan that
occur at one unit above the minimum.

In the hadronic tag analysis, we measure the branching fraction 
\begin{equation}
\mathcal{B}(\btn)=(1.8^{+0.9}_{-0.8}\pm 0.4 \pm 0.2) \times 10^{-4},
\label{eqn:bf}
\end{equation}
where the first error is statistical, the second is due to the background uncertainty, and the third is due 
to other systematic sources.
Taking into account the uncertainty on the expected background, as described above,  we obtain a significance 
of 2.2~$\sigma$.

In the semileptonic tag analysis, we measure the branching fraction
\begin{equation}
\mathcal{B}(\btn)=(0.9 \pm 0.6 \pm 0.1) \times 10^{-4},
\label{eqn:bf2}
\end{equation}
where the first error is statistical, and the second is due to the systematics uncertainties.

The combination of the two results yields: 
\begin{equation}
\mathcal{B}(\btn) = (1.2 \pm 0.4 \pm 0.3 \pm 0.2 ) \times 10^{-4},
\label{eqn:bfcombined}
\end{equation}
\noindent where the first uncertainty is statistical, the second is the uncertanty on the 
expected background and the third is from the other source of systematic effects.
The significance of the combined result is  2.6 $\sigma$ including the uncertainty on the expected 
background~(3.2 $\sigma$ if this uncertainty is not included). 


\begin{thebibliography}{999}
%
%
\bibitem{cc}
Charge-conjugate modes are implied throughout the paper.
\bibitem{c}
N. Cabibbo, Phys.\ Rev.\ Lett., \textbf{10} (1963) 531
\bibitem{km}
M. Kobayashi and T. Maskawa, Prog.\ Theor.\ Phys., \textbf{49} (1973) 652
\bibitem{higgs}
W. S. Hou, Phys.\ Rev.\ D  \textbf{48} (1993) 2342
\bibitem{Isidori2006pk}
G. Isidori and P. Paradisi, Phys.\ Lett. \textbf{B639} (2006) 499
\bibitem{Akeroyd:2003zr}
A. G. Akeroyd and S.  Recksiegel, J.\ Phys. \textbf{G29} (2003) 2311
\bibitem{belle}
Belle Collaboration, K. Ikado {\it et al.}, Phys. Rev. Lett. \textbf{97} (2006) 251802
\bibitem{taunusemilep}
\babar\ Collaboration, B.\ Aubert {\it et al.}, Phys. Rev D76:052002,2007  
\bibitem{taunuhad}
\babar\ Collaboration, B.\ Aubert {\it et al.}, submitted to Phys. Rev D, arXiv:0708.2260 [hep-ex]
\bibitem{thrust}
E.~Farhi, Phys. \ Rev.\ Lett. \textbf{39} (1977) 1587
\bibitem{pdg2004}
Particle Data Group, W. M. Yao  {\it et al.}, J. Phys. \textbf{G33} (2006) 1
\bibitem{prim1999}
J. Friedman and N. Fisher, Statistics and Computing \textbf{9} (1999) 123
\bibitem{crystalball}
M. J. Oreglia, SLAC-236 (1980)
\bibitem{arguspdf}
ARGUS Collaboration, H. Albrecht {\it et al.} Phys. Lett. \textbf{B185} (1987) 218
\bibitem{giunti}
C.\ Giunti, Phys.\ Rev.\ D \textbf{59} (1999) 113009
\end{thebibliography}
%

\end{document}